# Algorithm and Implementation of the Blog-Post Supervision Process

Kamanashis Biswas[1], Md. Liakat Ali[2], S.A.M. Harun[3]

**Abstract**— A web log or blog in short is a trendy way to share personal entries with others through website. A typical blog may consist of texts, images, audios and videos etc. Most of the blogs work as personal online diaries, while others may focus on specific interest such as photographs (photoblog), art (artblog), travel (tourblog), IT (techblog) etc. Another type of blogging called microblogging is also very well known now-a-days which contains very short posts. Like the developed countries, the users of blogs are gradually increasing in the developing countries e.g. Bangladesh. Due to the nature of open access to all users, some people misuse it to spread fake news to achieve individual or political goals. Some of them also post vulgar materials that make an embarrass situation for other bloggers. Even, sometimes it indulges the reputation of the victim. The only way to overcome this problem is to bring all the posts under supervision of the blog moderator. But it totally contradicts with blogging concepts. In this paper, we have implemented an algorithm that would help to prevent the offensive entries from being posted. These entries would go through a supervision process to justify themselves as legal posts. From the analysis of the result, we have shown that this approach can eliminate the chaotic situations in blogosphere at a great extent. Our experiment shows that about 90% of offensive posts can be detected and stopped from being published using this approach.

——————————  ◆  ——————————

## 1 INTRODUCTION

The old fashion of writing diaries is now changed to a new form known as blogging – a modern way of keeping diaries online. It is a platform where bloggers can share their daily lives, thoughts, problems, suggestions to others and blog readers can also post their feedbacks through comments or personal messages. As the users of computer and internet have increased dramatically all over the world, blogging has become a habit to many users. There are 184 million bloggers all over the world according to a 2008 study by Universal McCann [11]. Besides the developed countries, the people of the third world countries are also becoming more addicted to blogging day by day. For example, Indian blogosphere comprises more than 2 million blogs today. More than 50,000 people blog regularly in Bangladesh.

Like any other entity, blogs, though increasing popular day by day, are also facing some serious problems which have great impact on overall society. One of the major characteristics of blogs is that it provides open access to all. Anyone can register to a blog-site and he/she can post his personal journal anytime using his/her account. As there is no process of gate keeping on blogging, there is no check on what kind of content is being published by the bloggers.

The world of User Generated Content is full of material which puts negative impact on an individual, place, organization etc. which leads to defamation of somebody [1]. "The blogosphere has created an open, unfiltered megaphone for anyone with a computer and a modem, a striking characteristic of contemporary times is an unseemly incivility." says Robert Dilenschneider in his book 'power and influence'. This incivility includes making an untrue statement to a person or organization that damages the subject's reputation. Libel and Slander are two subcategories of defamation where first one is committed in printing media such as articles of a magazine or newspaper and the second one is in spoken form such as person-to-person or broadcast over a radio or television channel. Blogosphere very often suffers from defamation in a printed forum. A number of researches are done which include protection against spam blogs (Splog) [5, 10], instant blog updating and retrieving [4], providing fast news alert to the RSS feeds subscribers [7], technical (usability, reputation etc.) and social factors (community identification, attitudes towards blogging etc) related to acceptance of blog usages [3] . But no mechanism is proposed to overcome the above problem. In this paper, we have suggested a new approach which will work as a gate-keeper and will be able to minimize the number of offensive posts and comments.

## 2 COMMUNICATION MODEL OF BLOGS

### 2.1 TRADITIONAL BLOGGING SYSTEM

Robin Hamman has classified blogs into three categories. They are i) closed blogs, ii) blogs as conduit of information and iii) blog as participant in the conversation [2]. The most commonly used type of blog is 'blogs as conduit of information' as it includes a large number of participants. However, from the study of the underlying architecture of the current blogging system it is clear that most of the blogs provide open access to their users. That means that there is no supervision process for checking the misuse of the blogs. The following figure illustrates the current blogging system where bloggers directly write their entries using user interface module and submit it to the storage database. Immediate, it is published to the desired page without performing any verification of the posted contents. As a result, there is no gate-keeping mechanism followed in this process.

————————————
1. *Senior Lecturer, Daffodil International University, Department of CSE, Dhaka, Bangladesh.*
2. *Lecture, International Islamic University Chittagong, Department of EEE, Chittagong, Bangladesh.*
3. Senior Software Engineer, UNDP, *Dhaka, Bangladesh.*

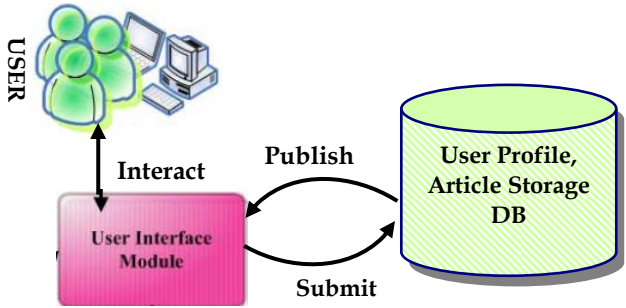

Fig. 1: Traditional Blogging System

## 2.2 PROPOSED BLOGGING SYTEM

This proposed system is better than the earlier model as each of the post must pass the verification phase. The following figure illustrates the process. In this system, every entry is assessed using an algorithm and if it achieves certain level of scores then will be published directly or directly with a notification. Otherwise it will wait in the queue for the approval of moderator or directly rejected on the basis of the score.

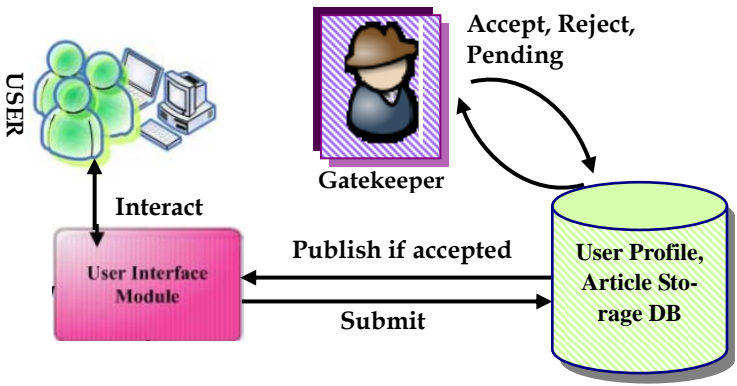

Fig. 2: Blogging under Supervision

## 3 A REAL BLOGGING SCENARIO

There are about 15 blogs in Bangladesh and most of them are developed for blogging in Bangla platform [12]. These blogs are categorized into community blogs, writers' forum, technical blogs etc. The first Bangla blog is somewhereinblog [8] which has around 40,000 registered accounts. About 10,000 bloggers visit, post, and interact on this site daily in average. At the beginning, this blog was open for everyone, i.e. anyone could register as a user. After account activation, he/she could entries directly. Unfortunately, some bloggers abused this opportunity by posting offensive materials. Some of these posts were to spread ideology of some forbidden groups. On the other hand, some posts and comments were made to spread fake news intentionally. During mid November, 2009, this misuse had been occurred so frequently that the site administrator imposed some restrictions. Now, the user's entry does not immediately appear in the home page after registration. It is only shown at his personal page. The blog administrator will observe new bloggers' activities and he/she will get access to the home page after getting safe blogger notification from administrator. There is also a rating system for the bloggers which indicates whether the blogger is in safe state or not. However, this chaos in somewhereinblog compelled many users to leave the site and at the same time a new blogging platform developed named as 'sachalayatan' [6] which is known as 'online writers' community'. This forum does not provide post directly to the home page. Users have to write their entries as guests and the entries have to wait for the approval of the moderator. Each user has to post as guest for long time before getting access to write as a regular member. Both the two blogs are following a supervision process which requires human interaction. Sometimes this takes long time and users get bored in blogging to these sites. This finally leads to decreasing users' interest.

## 4 BLOG-POST SUPERVISION PROCESS

We have proposed an algorithm that will check the blog objects which are submitted for publication. Here, we used a dictionary database which has two types of words and also a list of links of restricted sites. First type is slang word list which will be used to measure frequency of these words in a blog-post. Another type is demand-based word list. Anytime, new words can be added to or deleted from this list. Sometimes it is necessary to add some words which are not slang but required to control unusual events. For example, when a disaster (such as earthquake, fires [14] etc.) or political violence (such as revolt in PilKhana [14]) happens, then a number of blogs are posted immediately but most of them include wrong information about the incidents. These posts quickly spread rumor which may be harmful for the whole society. Demand base word list will help to protect the blogosphere from this type of chaotic situation.

### 4.1 BLOG-POST SUPERVISION ALGORITHM

*Procedure POST_SV ( Blog_Object)*

[ //Blog_Object may be Title, Post or Comments
  // Post or Comment includes body which may contain
    Texts, links of audio, video, documents or websites
  //Organize Blog_Object data for search, i.e. links will be
    searched first
]

*Do for each part of Blog_Object*
*While (True)*
*{*
     *//Continue till to end of selected module*
     *Integer  frequency_level, check*
     *//Initialize frequency_level and check to zero*

*Select an Item from Blog_Object data*
   *IF (Item is link and found in list)*
      *Reject the post, send a notification and set check to 1*
   *ELSE  IF  (Item is found in demand-based list)*
      *Keep it pending for the administrator's approval and set*
      *check to 1*
   *ELSE*
      *{*
        *For each Item in Database*
        *{*
           *IF (Found)*
              *Calculate the overall frequency of the Items*
        *} // End_For*

```
        }  // End_ELSE
      IF (check is not 1)
         IF (frequency_level > 40%)
               Reject the Post and send a notification
         ELSE IF (frequency_level is between 6-40%)
               Queue it for moderator's approval
         ELSE IF (frequency_level is between 1-5%)
               Publish it and send a notification
         ELSE
               Just publish it
   }  // End_While
```

## 5 IMPLEMENTATION AND ANALYSIS OF THE RESULT

The following figure shows the user interface that we have been developed using PHP for testing the efficiency of our proposed algorithm. We have performed our experiment in three phases. At the beginning, we have selected nine participants from Daffodil International University who are familiar with blogging. Five of them write entries frequently, two of them occasionally and other two only read blogs almost everyday. Before conducting the experiment, we have provided all the necessary information required to perform the tasks.

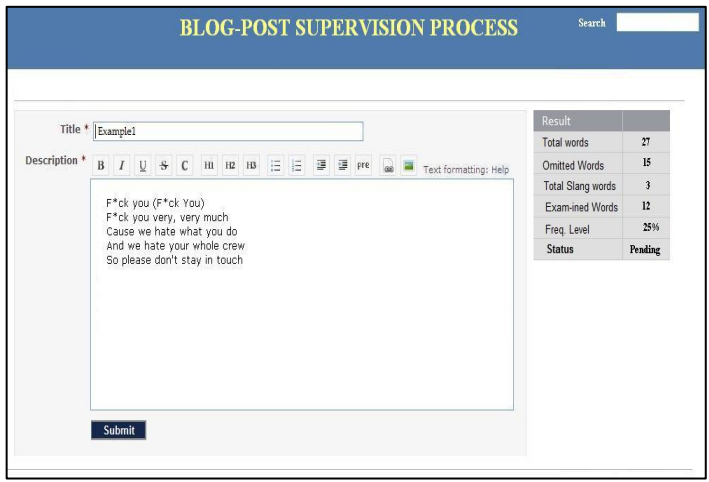

Fig. 3: User Interface Module

During first experiment, the participants are asked to write a short post including some vulgar links. After completion of the first phase, we have found that seven posts have been rejected out of nine. It means that our database contains the links of most commonly used offensive sites by the users.

Table 1: Outcome of Restricted Link Testing

| No. of Participants | Links Used | Links Matched | Rejected Post(s) | Published Post(s) |
|---|---|---|---|---|
| 9 | 15 | 12 | 7 | 2 |

Our second experiment has been done for testing the efficiency of demand-based post restriction mechanism. At this time, the participants are told to write a post about recent incident of Nimtoli fire [13] or any incident on fire. Before carrying out the test, we added three words (fire, nimtoli, burn) to the demand-based word-list. Six participants wrote on Nimtoli fire while other three described their own experiences on fires. The outcome shows that all nine posts are added to the waiting queue for the moderaotor's approval. This status indicates that selecting appropriate words can limit the number of entries regarding sensitive issues during urgent situation.

Table 2: Outcome of Demand-based Situation Testing

| Total Participants | Nimtoli Incident | Own Experiment | Pending Post(s) | Published Post(s) |
|---|---|---|---|---|
| 9 | 6 | 3 | 9 | 0 |

The last experiment is really tricky one. In this case, the participants are requested to write a post on their daily life and free to use slang words if they want. This time, we got very interesting output which is summarized in the next table. The algorithm considered all the words of given text and omitted the common terms such as auxiliary verbs (e.g. is, was, were, can, might, been etc.), prepositions (to, at, from, in, for etc.), articles (a, an, the), connective words (but, again, yet, because, still, however etc.) and pronouns (e.g. I, you, he, she, her, them etc.) for the simplification of the identifying frequency level. After filtering these types of words, the numbers of slang words are calculated using the dictionary database. Then, the frequency level of slang words is generated and on the basis of this score it is decided whether the post will be published or rejected or added to the waiting queue.

Table 3: Outcome of Post Text Testing

| Partic-ipants | Total words | Omitted Words | Total Slang words | Ex-amined Words | Freq. Level | Status |
|---|---|---|---|---|---|---|
| P1 | 45 | 18 | 3 | 27 | 11.11 | Pending |
| P2 | 205 | 135 | 8 | 70 | 11.43 | Pending |
| P3 | 318 | 133 | 12 | 185 | 6.49 | Pending |
| P4 | 56 | 33 | 10 | 23 | 43.48 | Rejected |
| P5 | 212 | 63 | 0 | 149 | 0 | Published |
| P6 | 315 | 158 | 2 | 157 | 1.27 | Published |
| P7 | 27 | 15 | 3 | 12 | 25 | Pending |
| P8 | 15 | 6 | 4 | 9 | 44.44 | Rejected |
| P9 | 159 | 51 | 0 | 108 | 0 | Published |

The above table shows that out of seven posts (including offensive contents), only one is published with notification, two posts are rejected and rest four are added to the waiting queue for moderator's approval. The overall result proves that the proposed algorithm for blog-post supervision is very efficient and able to eliminate the unpleasant posts at a great extent.

## 6 CONCLUSION

According to Andrew Sullivan [9], 'blogging is generating a new and quite essentially post modern idiom that's enabling users to express themselves in ways that have never been seen or understood before. Its truths are provisional, and its ethos collective and messy. Yet the interaction it enables between writer and reader is unprecedented, visceral, and sometimes brutal'. In fact, it is too hard to keep the blogosphere free from misuse. Only one way that can be done is check each and every part of the user generated contents. But it is totally contradictory with blog characteristics and also time consuming. Our pro-posed algorithm will play very effective role in this condition. It is very easy to find out the offensive posts and prohibit them from being post immediately. Though it is not possible to eliminate the current problem completely but still it will help to improve the chaotic situation of blogging environment. Our experiment shows that it is possible to reduce about 90% of offensive posts from being posted through our suggested mechanism. However, one of the big problems is that in some cases some vulgar words may be used in a post for different purpose (for example, tutorials). And there is a chance that this algorithm will treat the post as suspected one. Another possibility is that a post with very few vulgar words may be published due to low frequency level. However, except these problems, we can say that the algorithm is efficient enough to find out unpleasant entries very quickly and to handle the anomalous environment of the blogosphere.

### 6.1 FUTRE WORK

This algorithm can not check some of the posted or attached contents such as image or PDF files. It also cannot check the audio and video file contents for selecting a victim. There is scope to work in this area to improve the checking capability. Another property of the algorithm is that it is very simple. This can be improved using decision tree mechanism which will make a decision on the basis of the correlation of previous behavior of the user from the database and the frequency level. Finally, more experiments can be done to readjust the frequency level to achieve best output.

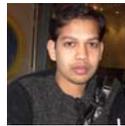
**Kamanashis Biswas**, born in 1982, post graduated from Blekinge Institute of Technology, Sweden in 2007. His field of specialization is on Security Engineering. At present, he is working as a Senior Lecturer in Daffodil International University, Dhaka, Bangladesh.

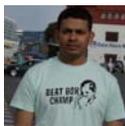
**Md. Liakat Ali,** born in 1981, post graduated from Blekinge Institute of Technology, Sweden 2007 in Security Engineering and 2008 in Telecommunication. Now he is working as a Lecturer in International Islamic University Chittagong, Bangladesh.

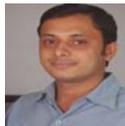
**S. A. M. Harun** is graduated from International Islamic University Chittagong. He is a programmer and ACM problem setter. Now he is working as senior software engineer at UNDP. His major area of interest is developing efficient algorithm.